\documentstyle[preprint,aps]{revtex}
\newcommand \beq{\begin{eqnarray}}
\newcommand \eeq{\end{eqnarray}}

\begin{document}
\draft
\preprint{LBL-37985, CERN-TH/95-304}

\title{$J/ \psi$ Suppression in an Equilibrating Parton Plasma}
\author{Xiao-Ming Xu$^{1,2}$, D. Kharzeev$^{3,4}$, H. Satz$^{3,4}$
and Xin-Nian Wang$^1$}
\address{$^1$Nuclear Science Division, Mailstop 70A-3307 \\
  Lawrence Berkeley National Laboratory, Berkeley, CA 94720}
\address{$^2$Theory Division, Shanghai Institute of Nuclear Research\\
  Chinese Academy of Sciences, P.O. Box 800204, Shanghai 201800, China}
\address{$^3$Theory Division, CERN, CH-1211 Geneva, Switzerland}
\address{$^4$Fakult\" at f\"ur Physik, Universit\" at Bielefeld\\
  D-33501 Bielefeld, Germany}

\date{November, 1995}
\maketitle

\begin{abstract}
\baselineskip=16pt

Short-distance QCD is employed to calculate the $J/\psi$
survival probability in an equilibrating parton gas, whose
evolution is governed by a set of master rate equations.
Partons in the early stage of high-energy nuclear collisions
may initially not be in equilibrium, but their average transverse
momentum is sufficiently high to break up a $Q\bar{Q}$ bound state.
Such a break-up during the evolution of the parton gas is shown to
cause a substantial $J/\psi$ suppression at both RHIC and LHC energies,
using realistic estimates of the initial parton densities.
The transverse momentum dependence of the suppression is also
shown to be sensitive to the initial conditions and the evolution
history of the parton plasma.

\end{abstract}

\baselineskip=18pt
\newpage
\section{Introduction}

It is generally believed, and confirmed by lattice QCD
calculations \cite{LATTICE}, that hadronic matter under
extreme conditions will form a plasma in which quarks and gluons
are no longer confined to individual hadrons and are in both
thermal and chemical equilibrium. To search for such a quark-gluon
plasma, it is proposed to study collisions of heavy nuclei
at extremely high energies. Recent work using models based on
perturbative QCD indeed shows that at high energy a dense partonic
system can be produced \cite{JBAM,KLL,HIJING,PCM,DTUNUC}.
Though it is not all clear if such partonic systems will
reach thermal and chemical equilibrium before hadronization
\cite{shuryak,kg92,bmtw,sinha,hhxw}, the partons, mainly gluons,
are certainly in a deconfined state at such high densities \cite{bmw}.

Many signals could arise from this deconfined state, such as
charm quark enhancement \cite{BMXW92,KG93,LG94} or enhanced
photon and dilepton production \cite{SX93,STLD}. Charm quarks,
for example, cannot be easily produced during the mixed and hadronic
phases of the dense matter, due to their large masses compared
to the temperature. They can be readily produced only
during the early stage of the evolution,
when partonic degrees of freedom are relevant. In this paper,
we will discuss how pre-equilibrium $J/\psi$ suppression can
be used to probe the early deconfined state of the partonic
system and the dynamics governing its evolution toward equilibrium.

$J/\psi$ suppression due to color screening has been proposed to
probe deconfinement \cite{ms}. This requires
that the interactions of $J/\psi$ with hadrons and
deconfined partons are different \cite{ks94}. Because
of its small size, a heavy quarkonium can probe the short-distance
properties of light hadrons. It is thus possible to make a parton-based
calculation of $J/\psi$-hadron cross section via an operator product
expansion method similar to that used in deeply-inelastic
lepton-hadron scatterings \cite{ks94,peskin,kaid}. The resulting
$J/\psi$-hadron cross section can be related to the distribution function
of gluons inside a hadron.
The energy dependence of the cross section near the
threshold of the break-up of a $J/\psi$ is determined by the
large $x$ behavior of the gluon distribution function, giving rise
to a very small break-up cross section at low energies. Only at very
high energies, this cross section will reach its
asymptotic value of a few mb.
  In other words, the  dissociation can only occur if the gluon
from the light hadron wave function is hard in the $J/\psi$'s rest frame,
i.e. its energy is high enough to overcome the binding energy threshold.
A hadron gas with temperature below 0.5 GeV certainly
cannot provide such energetic gluons to break up the $J/\psi$.
Therefore, a slow $J/\psi$ is very unlikely to be absorbed inside
a hadron gas of reasonable temperature \cite{ks94}. This
conclusion does not seem to be affected substantially by non-perturbative
effects, analyzed in Ref.\cite{kms}.

On the other hand, a deconfined partonic system contains much
harder gluons which can easily break up a $J/\psi$
\cite{ks94,ks95}.
A study of the energy dependence of the gluon-$J/\psi$ inelastic
cross section \cite{ks95} shows a strong peak just above
the break-up threshold of the gluon energy,
$\epsilon_0=2M_D-M_{J/\psi}$, where $M_{J/\psi}$ and $M_D$ are
the $J/\psi$ and D meson masses, respectively.
In the
pre-equilibrium stage, i.e., before the partons have reached
equilibrium, the average parton transverse momentum is sufficiently
large \cite{KEXW} to break up a $J/\psi$, provided the partons
are deconfined. The dissociation of $J/\psi$ will
continue during the whole equilibration process until the
effective temperature drops below a certain value or the
beginning of hadronization, whichever takes place first.
Therefore measurements of $J/\psi$
suppression can probe the deconfinement of the early partonic
system and shed light on the subsequent equilibration process,
provided that possible nuclear effects on the production of $Q{\bar Q}$
pairs and on pre-resonance charmonium states
are understood and taken into account.

In the following we will first calculate the thermal
gluon-$J/\psi$ dissociation cross section at different temperatures
and for different $J/\psi$ transverse momentum. We then follow
the evolution of an initially produced parton gas toward
equilibrium and calculate the resulting total survival probability
of a $J/\psi$ and its $P_T$-dependence.

\section{$J/\psi$ dissociation by gluons }

  The operator product expansion allows one to express the
hadron-$J/\psi$ inelastic cross section in terms of the
convolution of the gluon-$J/\psi$ dissociation cross section
with the gluon distribution inside the hadron \cite{ks94}.
The gluon-$J/\psi$ dissociation cross section is given by \cite{ks95}
\begin{equation}
  \sigma (q^0)= \frac{2\pi}{3}\left(\frac{32}{3}\right)^2
  \left(\frac{16\pi}{3g_s^2}\right)\frac{1}{m^2_Q}
  \frac{(q^0/\epsilon_0-1)^{3/2}}{(q^0/\epsilon_0)^5}\; , \label{eq1}
\end{equation}
where $g_s$ is the coupling constant of gluon and $c$ quark, $m_Q$
the $c$ quark mass, and $q^0$ the gluon energy in the $J/ \psi$ rest
frame;  its value must be larger than the $J/\psi$ binding energy
$\epsilon_0$. Since
for the tightly bound ground state of quarkonium
the binding force between the heavy quark and antiquark
is well approximated by the one-gluon-exchange Coulomb potential,
the $Q\bar{Q}$ bound state is hydrogen-like and the Coulomb relation
holds,
\begin{equation}
\epsilon_0 =\left({3g_s^2 \over 16\pi}\right)^2 m_Q \; . \label{eq2}
\end{equation}
The cross section thus can be rewritten as
\begin{equation}
  \sigma (q^0)=\frac{2\pi}{3}\left(\frac{32}{3}\right)^2
  \frac{1}{m_Q(\epsilon_0 m_Q)^{1/2}}
\frac{(q^0/\epsilon_0-1)^{3/2}}{(q^0/\epsilon_0)^5}\; . \label{eq3}
\end{equation}

As shown in Monte Carlo simulations \cite{KEXW}, the parton
density in the early stage of high-energy heavy-ion collisions
has an approximate Bjorken-type \cite{bj83} scaling behavior. We
will only consider $J/\psi$ suppression in the central rapidity
region ($y_{J/\psi}\simeq 0$). In this case, the $J/ \psi$ will move
in the transverse direction with a four-velocity
\begin{equation}
u=(M_T, \vec{P_T}, 0)/M_{J/\psi}, \label{eq4}
\end{equation}
where $M_T=\sqrt{P_T^2+M^2_{J/ \psi}}$ is defined as the $J/\psi$'s
transverse mass. A gluon with a four-momentum $k=(k^0,\vec{k})$
in the rest frame of the parton gas has an energy $q^0=k\cdot u$
in the rest frame of the $J/\psi$. The thermal
gluon-$J/\psi$ dissociation cross section is then defined as
\begin{equation}
\langle v_{\rm rel} \sigma (k \cdot u)\rangle_k
= \frac{\int d^3k v_{\rm rel} \sigma (k
\cdot u) f(k^0;T)}{\int d^3k f(k^0;T)} \; , \label{eq5}
\end{equation}
where the gluon distribution in the rest frame of the
parton gas is defined as
\begin{equation}
  f(k^0;T)=\frac{\lambda_g}{e^{k^0/T}-1} \label{eq6}
\end{equation}
with $\lambda_g\leq 1$ specifying the deviation of the system from
chemical equilibrium. The relative velocity $v_{\rm rel}$ between
the $J/\psi$ and a gluon is
\begin{equation}
  v_{\rm rel}=\frac{P_{J/\psi}\cdot k}{k^0M_T}
  =1-\frac{\vec{k}\cdot\vec{P}_T}{k^0M_T}\; . \label{eq7}
\end{equation}
Changing the variable to the gluon momentum, $q=(q^0,\vec{q})$, in
the rest frame of the $J/\psi$, the integral in the numerator
of Eq.~(\ref{eq5}) can be rewritten as
\begin{equation}
  \int d^3q \frac{M_{J/\psi}}{M_T}\sigma(q^0) f(k^0;T), \label{eq8}
\end{equation}
where
\begin{equation}
  k^0=(q^0M_T+\vec{q}\cdot\vec{P}_T)/M_{J/\psi} \; . \label{eq9}
\end{equation}
One can carry out the integral in the denominator,
$\int d^3k f(k^0;T)=8\pi \zeta(3)\lambda_g T^3$, and the angular
part in the numerator, to get
\begin{equation}
  \langle v_{\rm rel}\sigma(k\cdot u)\rangle_k=
(\frac{8}{3})^3\frac{\pi}{\zeta(3)}\frac{M_{J/\psi}^2}{P_TM_T T^3}
(\frac{\epsilon_0}{m_Q})^{3/2}\sum_{n=1}^{\infty}T_n
\int_1^{\infty} dx\frac{(x-1)^{3/2}}{x^4}(e^{-a_n^-x}-e^{-a_n^+x})\; ,
\label{eq10}
\end{equation}
with $T_n=T/n$ and
\begin{equation}
  a_n^{\pm}=\frac{\epsilon_0}{T_n}\frac{M_T\pm P_T}{M_{J/\psi}}
  \; . \label{eq11}
\end{equation}

In order to understand the temperature and $P_T$ dependence of the
thermal gluon-$J/\psi$ dissociation cross section,
we first plot in Fig.~1 the cross section $\sigma(q^0)$ of
Eq.~(\ref{eq3}) as a function of the gluon energy in the $J/\psi$' rest frame.
It decreases strongly toward the threshold and is broadly
peaked around $q^0=10\epsilon_0/7=0.92$ GeV, with a maximum value
of about 3 mb. Low-momentum gluons have neither the resolution to
distinguish the heavy constituent quarks nor the energy to
excite them to the continuum. On the other end, high-momentum
gluons also have small cross section with a $J/\psi$ since they
cannot see the large size.

We can also express the cross section as a function of the
center-of-mass energy of gluons and the $J/\psi$,
$\sigma(q^0)=\sigma(s/2M_{J/\psi}-M_{J/\psi}/2)$, where
$s=(k+P_{J/\psi})^2$. One can thus translate the energy
dependence in Fig.~1 into temperature and $P_T$ dependences
after thermal average, since the thermally averaged $<s>$
is proportional to both $P_T$ and temperature $T$.
In Fig.~2 we plot the thermally averaged gluon-$J/\psi$ dissociation
cross section as a function of temperature for
different values of the $J/\psi$'s transverse momentum $P_T$.
We observe the same kind of peak structure, with a decreased
maximum value due to the thermal average. The position of the
peak also shifts to smaller values of $T$ when $P_T$ is
increased, corresponding to a fixed value of the averaged
center-of-mass energy $\langle s\rangle$. A similar behavior is
expected if one plots the thermal
cross section as a function of $P_T$ at different temperatures, as done
in Fig.~3. However, in this case, the peak simply disappears
at high enough temperatures, because the averaged $\langle s\rangle$
will be above the threshold value even for $P_T=0$.
These features will have considerable consequences for
the survival probability of a $J/\psi$ in an equilibrating
parton gas, especially the $P_T$ dependence.
We should also mention that the use of the Bose-Einstein
distribution function has an effect of about 20\% on the
thermal cross section, compared to that
obtained with a Boltzmann distribution \cite{ks95}.

\section{ $J/\psi$ suppression in an equilibrating parton gas}

Using the thermal cross section just obtained, we can now
calculate the survival probability of $J/ \psi$ in an
equilibrating parton plasma. In this paper, we will neglect the
transverse expansion and consider only longitudinal expansion.
We will also only consider $J/\psi$ suppression in the central
rapidity region. A $J/\psi$ produced at point $\vec{r}$
with velocity $\vec{v}$ in the transverse
direction will travel a distance
\begin{equation}
  d=-r\cos\phi + \sqrt{R_A^2-r^2(1-\cos^2\phi)} \label{eq12}
\end{equation}
in the time interval $t_{\psi}=M_T d/P_T$ before it escapes from a
gluon gas of transverse extension $R_A$; here
$\cos\phi= \hat{\vec{v}}\cdot\hat{\vec{r}}$. Suppose the
system evolves in a deconfined state until the temperature
drops below a certain value, which we assume to be 200 MeV.
The total amount of time the $J/\psi$ remains
inside a deconfined parton gas is the smaller one of the two times
$t_{\psi}$ and $t_f$,
the life-time of the parton gas. Assume that the
initial production rate of the $J/\psi$ is proportional to the number
of binary nucleon-nucleon interactions at impact-parameter $r$,
$N_A(r)=A^2(1-r^2/R_A^2)/2\pi R_A^2$. The survival
probability of the $J/\psi$
averaged over its initial position
and direction in an equilibrating parton gas is
\begin{equation}
  S(P_T)=\frac{\int d^2r (R_A^2-r^2) \exp{[-\int^{t_{\rm min}}_0 d\tau
      n_g(\tau)\langle v_{\rm rel}\sigma(k\cdot u)\rangle_k]}}
  {\int d^2r (R_A^2-r^2)} \; , \label{eq13}
\end{equation}
where
\begin{equation}
  t_{\rm min}=\min (t_{\psi},t_f) \; , \label{eq14}
\end{equation}
and $n_g(\tau)$ is the gluon number density at a given time  $\tau$.

In Eq.~(\ref{eq13}), both the gluon number density $n_g(\tau)$
and the thermal cross section $\langle v_{\rm rel}\sigma(k\cdot u)\rangle_k$
depend on the temperature, which in turn is a function of time.
In addition, $n_g$ is proportional to gluon fugacity which also
evolves with time. To evaluate the survival probability we need
to know the entire evolution history of the parton system.
Roughly speaking, one can divide this history
into two stages: (1) First there is kinetic thermalization,
mainly through elastic scatterings and expansion.
The kinematic separation of free-streaming partons gives us
$\tau_0 \sim 0.5$-0.7 fm/$c$
as an estimate of the time
when local isotropy in momentum distribution is reached
\cite{bmtw,KEXW}.
(2) The parton gas now further evolves toward chemical
equilibrium through parton proliferation and gluon fusion; it does so
until hadronization or freeze-out,  whichever happens first.
This evolution can be determined by a set of master rate
equations which give us the time dependence of the temperature
and fugacities.

In this paper we will not address the question of pre-resonance
$J/\psi$ suppression, which has been discussed and shown to be
responsible for the $J/\psi$ suppression observed in $p-A$ and
$S-U$ collisions \cite{gh92,boresk,cywong,ksc}.
We will here consider the suppression of fully formed physical
$J/\psi$ states as it should take place
if nuclear collision produces dense
partonic system.

Following Ref.~\cite{bmtw}, we characterize the non-equilibrium
of the system by gluon and quark fugacities which are less
than unity. The dominant reactions leading to chemical
equilibrium are assumed to be the following two processes:
\begin{equation}
  gg \leftrightarrow ggg, \;\;\;\; gg \leftrightarrow q \bar{q}
  \; . \label{eq15}
\end{equation}
Assuming that elastic parton scatterings are sufficiently rapid
to maintain local thermal equilibrium, the evolution of the
parton densities can be given by the master rate equations.
Combining these master equations together with one-dimensional
hydrodynamic equation, one can get the following set of
equations \cite{bmtw}:
\begin{eqnarray}
  \frac{\dot{\lambda}_g}{\lambda_g} +3 \frac{\dot{T}}{T}
    +\frac{1}{\tau} & = & R_3(1-\lambda_g) -
    2R_2(1-\frac{\lambda^2_q}{\lambda_g^2}), \\
    \frac{\dot{\lambda}_q}{\lambda_q} + 3\frac{\dot{T}}{T}
    +\frac{1}{\tau} & = & R_2\frac{a_1}{b_1}(\frac{\lambda_g}{\lambda_q}
    - \frac{\lambda_q}{\lambda_g}), \\
    (\lambda_g + \frac{b_2}{a_2}\lambda_q)^{3/4}T^3 \tau
    & = & {\rm const.},
\end{eqnarray}
where $a_1=16\zeta (3)/\pi^2\approx 1.95$, $a_2=8\pi^2/15\approx 5.26$,
$b_1=9\zeta (3)N_f/\pi^2\approx 2.20$ and $b_2=7\pi^2N_f/20 \approx 6.9$.
The density and velocity weighted reaction rates
\begin{equation}
R_3 = \frac{1}{ 2} \langle\sigma_{gg\rightarrow ggg}v\rangle n_g, \quad
R_2 = \frac{1}{2} \langle\sigma_{gg\rightarrow q\bar{q}}v\rangle n_g,
\end{equation}
can be found in Refs.~\cite{bmtw,BMXW92}. After taking into account
parton screening \cite{bmw} and Landau-Pomeranchuck-Migdal
effect in induced gluon radiation   \cite{wgp,baier}  ,
$R_3/T$ and $R_2/T$
are found to be functions of only $\lambda_g$.
Solving the above rate equations as shown in Ref.~\cite{bmtw},
one finds that the parton gas cools considerably
faster than predicted by Bjorken's scaling
solution  $(T^3\tau$ = const.), because the production of
additional partons approaching the chemical equilibrium state
consumes an appreciable amount of energy. The accelerated cooling,
in turn, slows down the chemical equilibration process.
For the initial conditions given by HIJING Monte Carlo
simulations \cite{HIJING}, at RHIC energy
the parton system can hardly reach its equilibrium state,
since the effective temperature here
drops below $T_c \approx 200$ MeV in a time of some
1-2 fm/$c$. At LHC energy, however, the parton
gas comes very close to equilibrium, since the system
may exist in a deconfined state for as long as 4-5 fm/$c$.
In the following we will use the numerical results for the
time evolution of the temperature and fugacities obtained from the
above master equations to calculate the survival probability
of a $J/\psi$ in such an equilibrating parton system.

Shown in Fig.~4(a) are the $J/\psi$ survival
probabilities in the deconfined and equilibrating parton
plasma at RHIC and LHC energies with initial conditions
given by HIJING Monte Carlo simulations \cite{HIJING},
denoted as set 1 in Table \ref{table1}. We  find that there is stronger
$J/\psi$ suppression at LHC than at RHIC energy, due both to the
higher initial parton densities and longer life-time of the
parton plasma. The increase of the survival probabilities with
$J/\psi$'s transverse momentum is a consequence of the
decrease of the thermal cross section with increasing $P_T$
at high temperatures, as shown in Fig.~3, and the shorter time
spent by a higher-$P_T$ $J/\psi$ inside the parton plasma, an
effect first considered in Ref.\cite{kp}.
$J/\psi$ dissociation during the late stage of the evolution
should have a peak and increase a little with $P_T$ when
the temperature drops below 0.3 GeV, as illustrated in Fig.~3.
This behavior flattens the total survival
probability at small values of $P_T$ when we integrate over
the entire history of the evolution from high initial temperatures.
For a parton system with a low initial temperature (below 300 MeV),
the $P_T$ dependence of the survival probability should be
even flatter. One can therefore use the $P_T$ dependence to
shed light on the initial temperature and the evolution history
of the system.

To demonstrate the effects of the chemical non-equilibrium in
the initial system, we also show in Fig.~4(b)
the survival probabilities in an ideal parton gas with the same
initial temperatures as before, but with full chemical
equilibrium (unit initial fugacities). In this case, the temperature
simply decreases like $T(\tau)=T(\tau_0)({\tau_0/\tau})^{1/3}$.
We see that the $J/\psi$ is now much more suppressed than in the case
of an equilibrating parton plasma, because of both the higher
parton density and the longer life time of the system.

Since there is considerable uncertainty in the estimate of the
initial parton production by HIJING Monte Carlo simulations,
as discussed in Ref.~\cite{bmtw}, we would like to test the sensitivity
of $J/\psi$ suppression to this. We therefore
multiply initial parton number densities by a factor of 4,
thus increasing initial parton fugacities. We denote such
initial conditions as set 2 in Table~\ref{table1}. If the uncertainties
in initial conditions are caused by the soft parton
production from the color mean fields, the initial effective
temperature will decrease. Therefore, we can alternatively
increase the initial parton density by a factor of 4 and
at the same time decrease $T_0$ to 0.4 and 0.72 GeV at RHIC and
LHC energies, respectively. This leads to higher initial fugacities,
listed as set 3 in Table~\ref{table1}.
The corresponding survival probabilities
calculated with these two sets of initial conditions are shown
in Fig.~5. We can see that the $J/\psi$ suppression is much stronger
if the initial parton densities are higher. Comparing the
solid and dashed lines, however, shows that the $J/\psi$
suppression is less sensitive to the variation of the initial
temperature and fugacities as far as the parton densities
are fixed.

\section{Conclusions}

To summarize, we have used the  cross section of
$J/ \psi$ dissociation
by gluons to calculate the $J/\psi$ suppression in an
equilibrating parton gas produced in high-energy nuclear
collisions. The large average momentum in the hot gluon
gas enables gluons to break up the $J/\psi$, while
hadron matter at reasonable temperature does not provide sufficiently
hard gluons.   We find a substantial $J/\psi$
suppression in such a non-equilibrium partonic medium; however,
it is
smaller than that in a fully equilibrated parton plasma.
In particular, in an equilibrating plasma the behavior of the $J/\psi$-gluon
cross section at high gluon momenta reduces the $J/\psi$ suppression
at large $P_T$.

In addition to the $J/\psi$ dissociation during the equilibration
of the parton plasma, there are other possible sources of suppression
for the actually observed $J/\psi$'s. As already noted, nuclear
modifications of the $Q{\bar Q}$ production process e.g., through
modified gluon distributions in a nucleus \cite{MQ86,EQW94}, multiple
scattering accompanied by energy loss \cite{gm92}, or a suppression of
the nascent $J/\psi$ before it forms an actual physical resonance
\cite{ksc} must be taken into account. Such effects would
cause $J/\psi$ suppression in addition to what we have obtained
from the equilibrating parton plasma and modify the transverse
momentum dependence of $J/\psi$ suppression \cite{ggj}.
Moreover, interactions between gluons and $c\bar{c}$ bound states
before the kinetic thermalization of the partons could also lead
to a substantial $J/\psi$ suppression; this would depend
on the time needed to achieve local momentum isotropy.
On the other hand, gluon fusion could also result in
$J/\psi$ production during the evolution of the parton system,
similar to the enhancement of open charm. Although
studies of pre-equilibrium open charm production
indicate \cite{BMXW92,LG94} that $J/\psi$ production during the parton
evolution is not significant compared to primary production, a
consistent study of $J/\psi$ suppression should include
the pre-equilibrium production in a form of a master rate
equation.

\section*{ACKNOWLEDGEMENTS}
X.X thanks the Nuclear Theory Group at LBL Berkeley for their
hospitality during his visit, when this work was carried out.
D.K, H.S and X.-N.W. thank K.~J.~Eskola for stimulating discussions.
This work was supported by the U.S. Department of Energy
under Contract No. DE-AC03-76SF00098 and by the German Research
Ministry (BMBW) under the contract 06 BI 721.

\begin{table}
\begin{center}
\begin{tabular}{lllllll}
\mbox{} & RHIC(1) & LHC(1) & RHIC(2) & LHC(2) & RHIC(3) & LHC(3) \\\hline
$T$(GeV)  & 0.55 & 0.82 & 0.55 & 0.82 & 0.4  & 0.72   \\
$\lambda_g$& 0.05 & 0.124& 0.2  & 0.496&0.53 & 0.761  \\
$\lambda_q$&0.008 & 0.02 & 0.032& 0.08 &0.083&0.118  \\
$n_g$(fm$^-3$)& 2.15 & 18 &8.6 &72 &8.6 &72  \\
$n_q$(fm$^-3$)&0.19 &1.573&0.76&6.29&0.76&6.29 \\
\end{tabular}
\caption{Different sets of initial conditions of the temperature,
  fugacities and parton number densities at $\tau_0=0.7$fm/c
  for RHIC and $\tau_0=0.5$fm/c for LHC.} \label{table1}
\end{center}
\end{table}

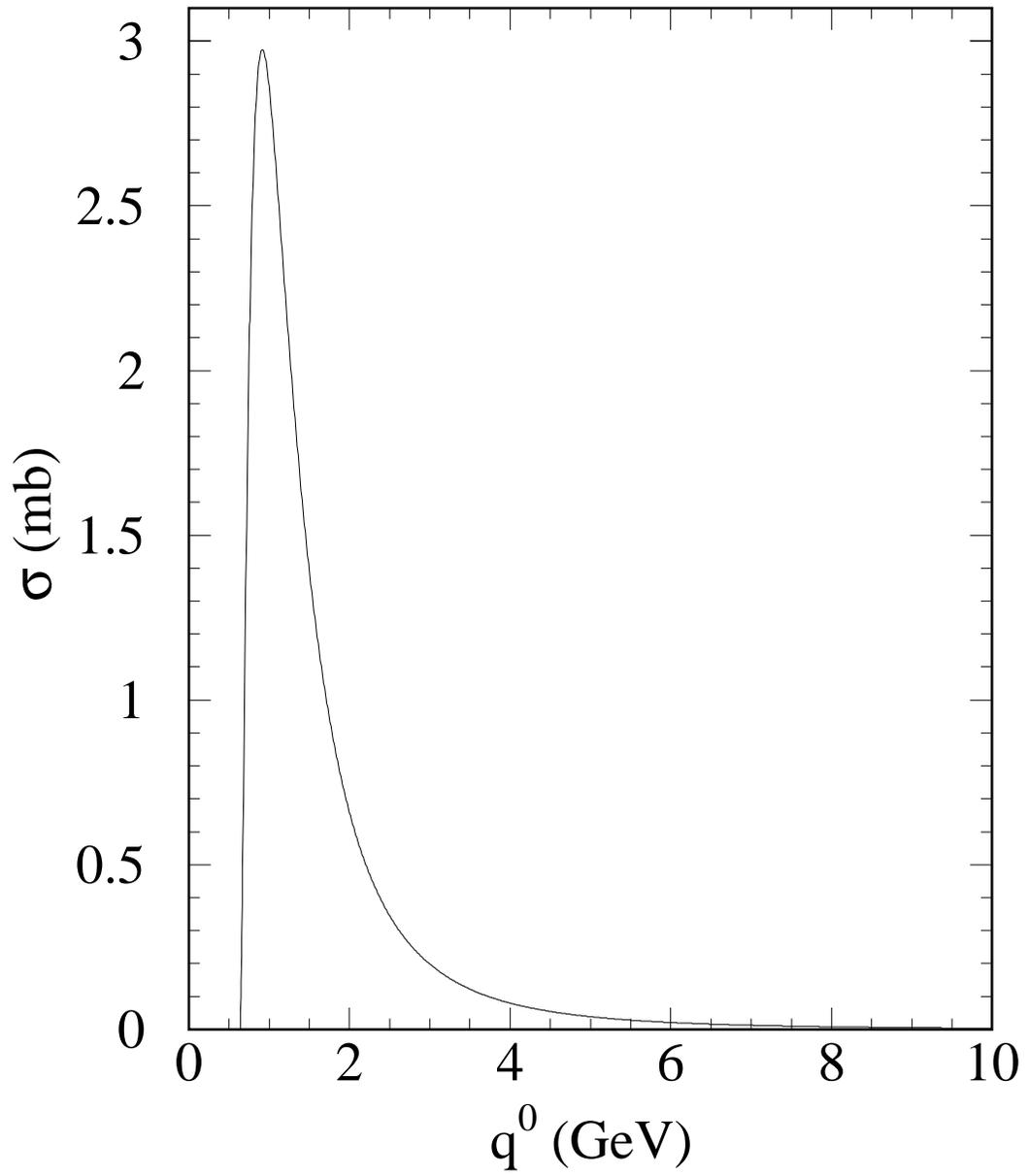
\begin{figure}
\caption{Gluon-$J/ \psi$ dissociation cross section as a function
  of the gluon energy $q^0$ in the rest frame of the $J/\psi$.}
\end{figure}

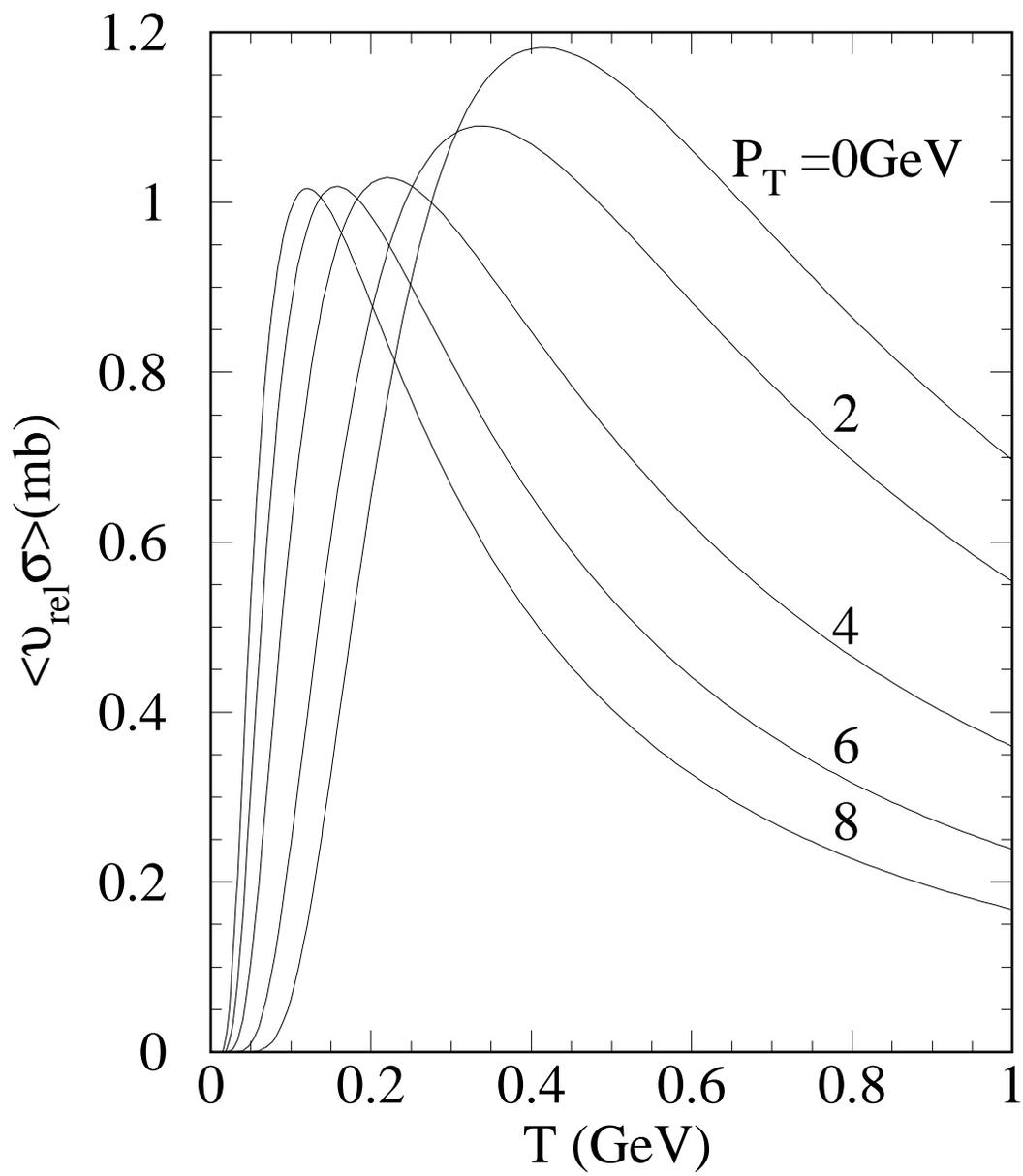
\begin{figure}
\caption{The thermal-averaged gluon-$J/\psi$ dissociation cross
  section $<v_{\rm rel}\sigma>$ as a function of the temperature
  at different transverse momenta $P_T$.}
\end{figure}

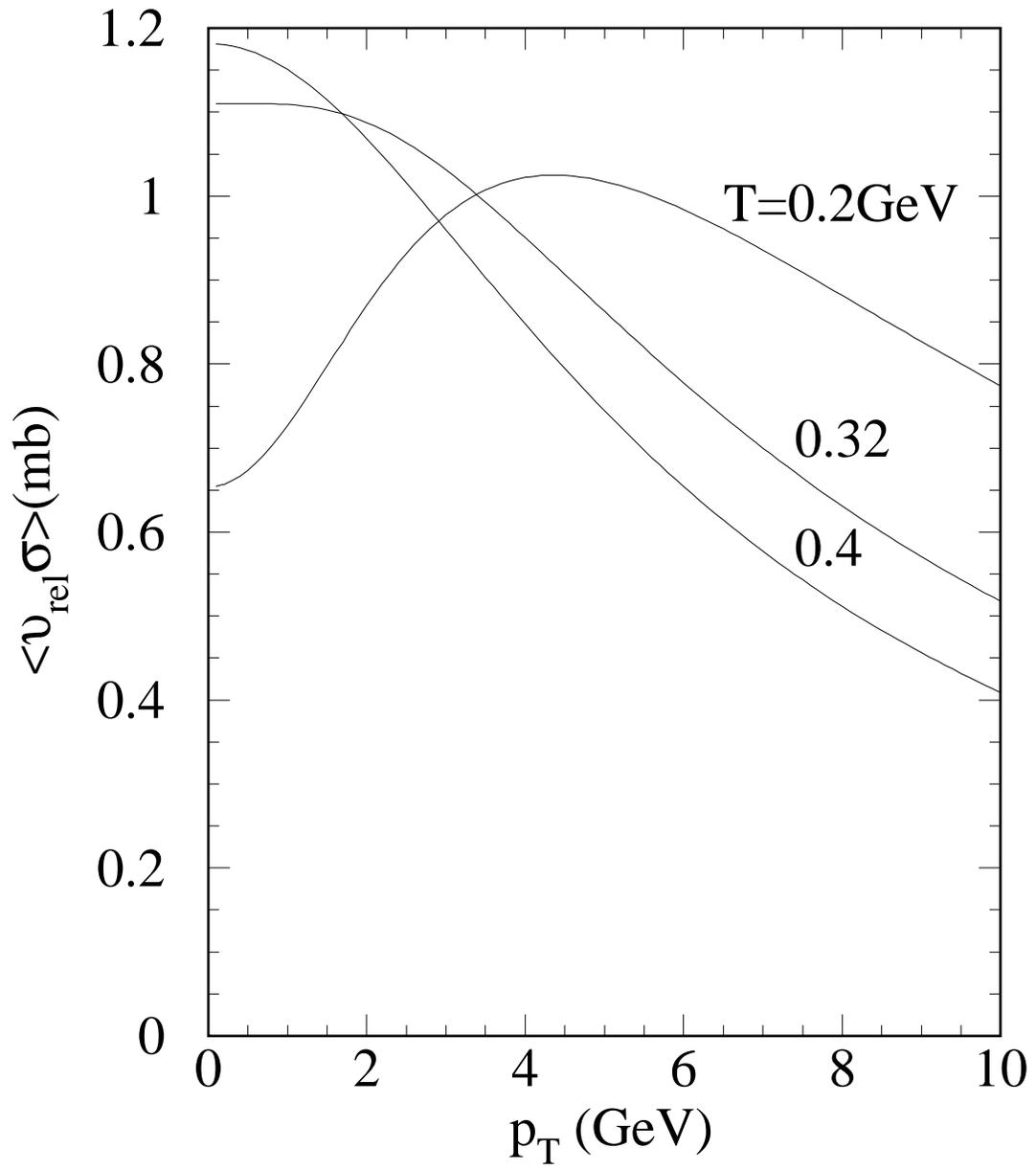
\begin{figure}
\caption{The thermal-averaged gluon-$J/\psi$ dissociation cross
  section $<v_{\rm rel}\sigma>$ as a function of the transverse
  momentum $P_T$ at different temperatures.}
\end{figure}

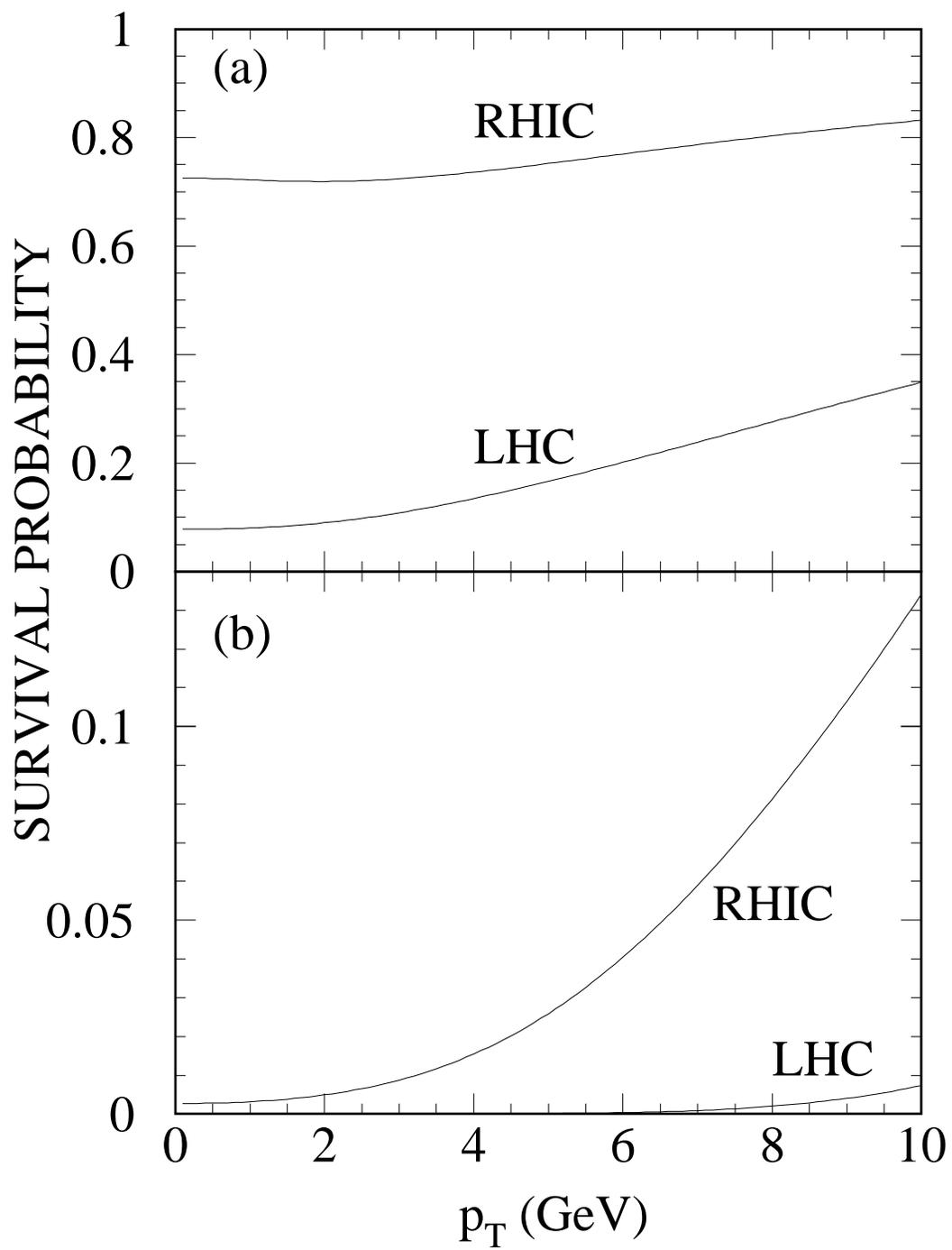
\begin{figure}
\caption{ (a) The survival probability of $J/\psi$ in an equilibrating parton
  plasma at RHIC and LHC energies with initial conditions given as
  set 1 in Table~\protect\ref{table1}, (b) and for an initially equilibrated
  plasma at the same temperatures.}
\end{figure}

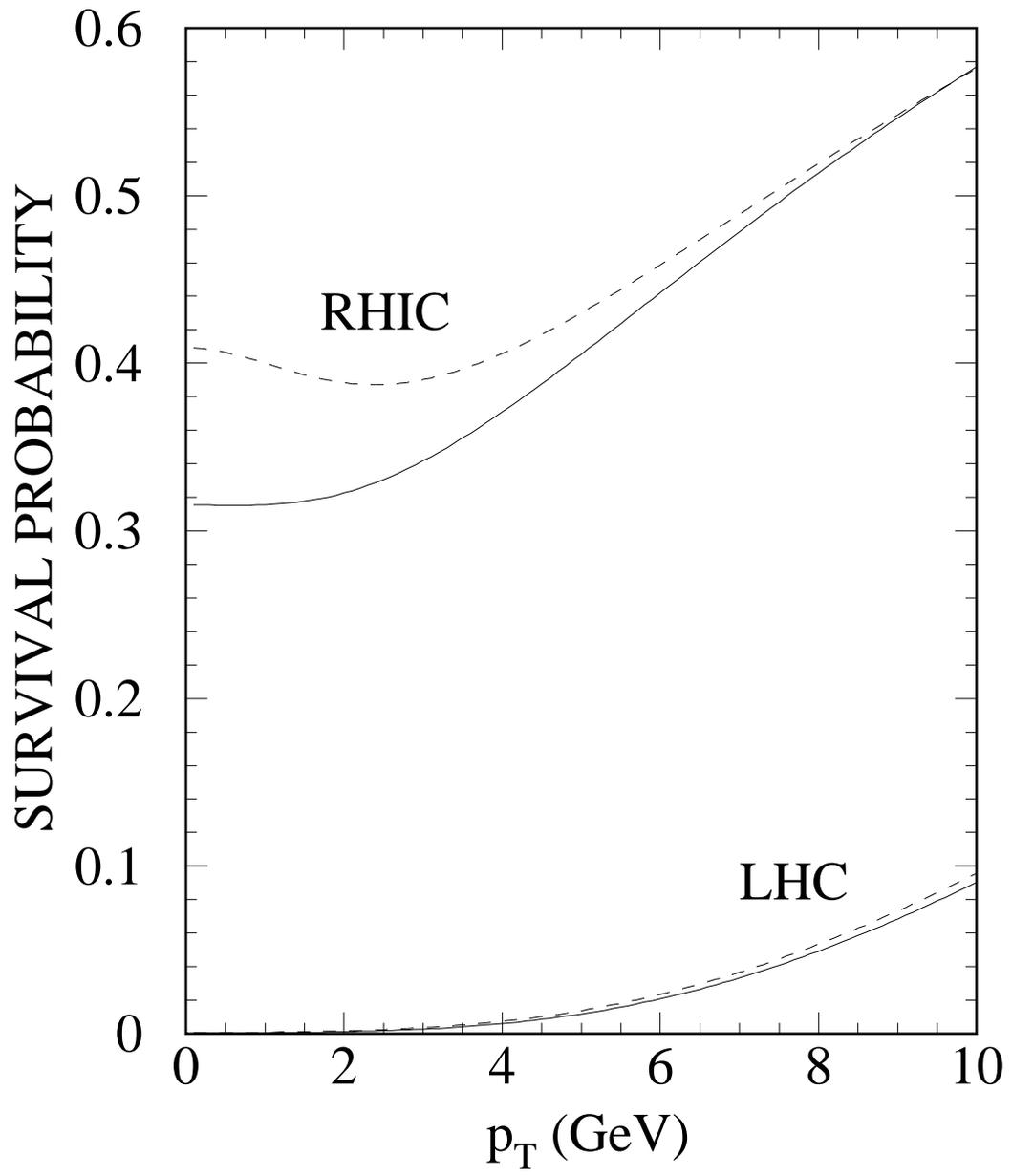
\begin{figure}
\caption{The survival probability of $J/\psi$ in an equilibrating parton
  plasma with initial conditions given as set 2 (solid) and set 3 (dashed)
  in Table~\protect\ref{table1}.}
\end{figure}

\end{document}